\def\year{2021}\relax
\documentclass[letterpaper]{article} 
\usepackage{aaai21}  
\usepackage{times}  
\usepackage{helvet} 
\usepackage{courier}  
\usepackage[hyphens]{url}  
\usepackage{graphicx} 
\usepackage{tabularx}%
\usepackage{multirow}%
\usepackage{natbib}  
\usepackage{caption} 
\title{Morphological Matrices as a Tool for Crowdsourced Ideation}
\author{
    Jonas Oppenlaender 
    \\
}
\affiliations{
    University of Oulu\\
    jonas.oppenlaender@oulu.fi
}

\begin{document}

\maketitle

\begin{abstract}%
Designing a novel product is a difficult task not well suited for non-expert crowd workers. In this work-in-progress paper, we first motivate why the design of \textit{persuasive} products is an interesting context for studying creativity and the \textit{creative leap}.
We then present a pilot study on the crowdsourced design of persuasive products. The pilot study motivated our subsequent feasibility study on the use of morphological matrices as a tool for crowdsourced ideation and product design. Given the morphological matrix, workers were able to come up with valid and significantly more relevant ideas for novel persuasive products.
\end{abstract}%

\section{Background: The Magical Creative Leap} 
\label{introduction}%

\noindent
Designing a novel product is a difficult creative task not well suited for microtask crowdsourcing platforms.
The challenges in enabling complex creative work via crowdsourcing have been underexplored, and for crowdsourcing to expand its territories beyond simple human perception and data labeling tasks,
more research is needed.
Crowdsourcing not only holds opportunities for supplying feedback for creative work \cite{CC19DoctoralConsortium}, 
but also to support 
design activities with
\textit{crowd-powered creativity support systems} \cite{DC2S2,CreativitySupport}.
Consider the example of designing a 
\textit{persuasive product}. Persuasive products non-intrusively employ design
friction ``to reinforce, change or shape attitudes or behaviors [...] 
without using coercion or deception''~\cite{design-principles}.
    An example 
    is \textit{Keymoment} by \citet{keymoment}. 
    Keymoment is a key holder that will persuade its users to use the bicycle (healthy option) by dropping the bicycle key to the floor if one reaches for the car key (less healthy option).

Imagining a mechanism for a novel persuasive product is a difficult design task. 
This makes the design of persuasive products an interesting context for studying creativity and in particular the stage in the creative process which we call the \textit{creative leap}.
%
%
According to Wallas' four-stage model of the creative process \cite{WALLAS}, the stage in the creative process related to
    the creative leap
    is \textit{illumination}.
\citet{kolko.pdf} referred to this step in the design process
as a ``frustrating part of product development'' which may appear ``magical when encountered  in  professional practice.''
\section{Pilot Study}%

We asked workers (from the United States and the United Kingdom with greater than 100 Human Intelligence Tasks (HITs) completed and greater than 95\% HIT approval rate, US\$ 1 per HIT) on Amazon Mechanical Turk (MTurk) to 
come up with the design for a novel persuasive product.
The HIT included descriptions and examples of four design principles of ``pleasurable troublemakers'' \cite{Laschke2014,3461778.3462079.pdf}: situatedness, alternative choice, freedom, meaning-making.
For each principle, workers were to come up with one idea of how a product feature could address the principle.
Next, workers were to combine the four features into one cohesive design of a persuasive product.
Workers were instructed not to use a search engine.
Tasks were released in small batches during the day.

%
%

\subsubsection{Pilot study results}
    Twelve workers (8~male, 4~female,
        aged 23--61, mean age 38.6 years)
        elected to complete the task.
    Tasks had to be rejected in all 12 cases (100\%).
    Workers either tried to game the task (e.g. by copy-pasting parts of the task instructions) or
    clearly spent no own cognitive effort on completing the task (e.g. by responding with texts copied from websites or search engine results).

The pilot study surfaced recurring issues with subjective creative work gathered in open-ended tasks on MTurk reflected in prior literature \cite{kittur2008,p1631-gadiraju.pdf,CHI2020,THESIS}. 
    ``Malicious workers'' exploit and game survey tasks for financial gains and a large part of the open-ended responses are untrustworthy~\cite{p1631-gadiraju.pdf}. 

\section{Morphological Matrices to Structure Ideation}%

How can we elicit novel designs for persuasive products from non-expert crowd workers?
The crowd workers will need to receive methodological support to complete this difficult task.
%
Rubrics \cite{Posts_paper_3.pdf}, scaffolds \cite{3450741.3465261.pdf,3357236.3395480.pdf}, and guiding questions \cite{paper138.pdf} have proven efficient to structure the thought processes of crowd workers and to scaffold idea generation.
If we could reduce the number of open-ended items in the task, we may be able to ``minimize'' the creative leap and reduce cognitive burden on workers.

\textit{Morphological matrices} (MM; also called morphological charts, morphological boxes, or concept combination tables) may be a way to achieve this goal.
MM are a method used in New Product Development (NPD) to generate ideas by combining individual product features into new product designs~\cite{productdev,pahlbeitz2007}.
MM apply the principle of functional decomposition.
The columns in the matrix represent a collection of functional solutions to sub-problems.
A solution to the overall problem is selected by combining one solution from each column.

Given the MM, crowd workers can ideate a product by selecting different features (one from each column).
    Self-selection has proven effective in not limiting a worker's creativity \cite{MUM}.
The selection is then synthesized into a product design. This would require only one open-ended item for describing how the whole product would function given the selected features.
Thus, the number of open-ended items is reduced and the cognitive effort to solve the task is minimized.
\section{Feasibility Study}

\begin{figure}[t]%
\centering%
\noindent\fbox{%
    \parbox{0.95\columnwidth}{%
\includegraphics[width=0.95\columnwidth]{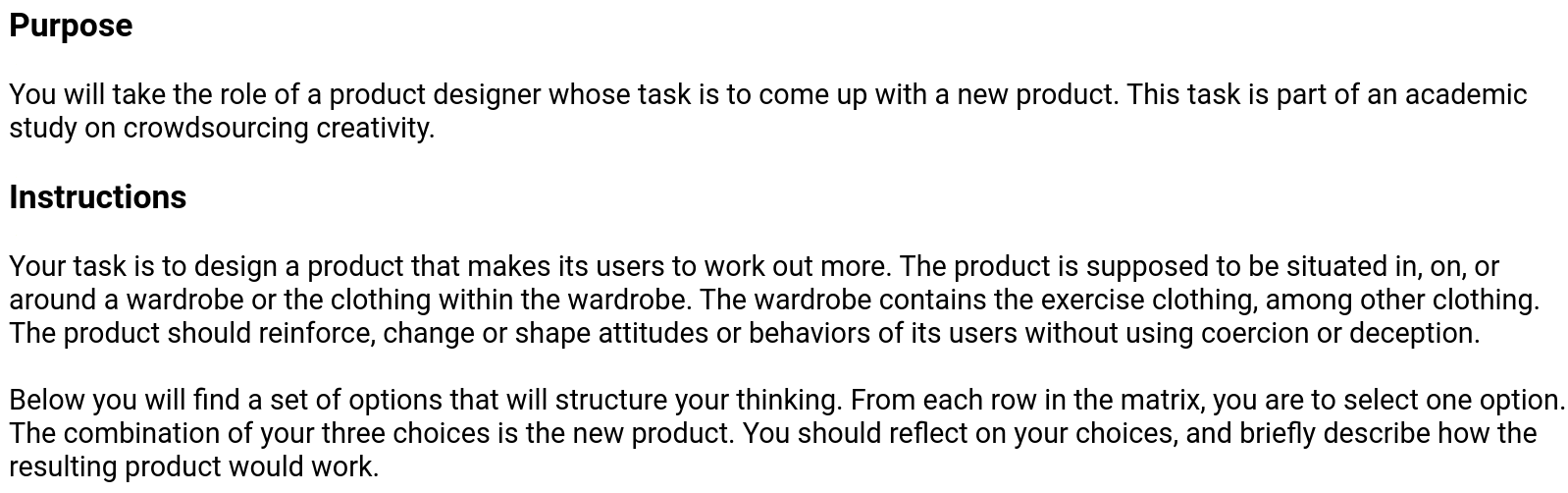}
\includegraphics[width=0.95\columnwidth]{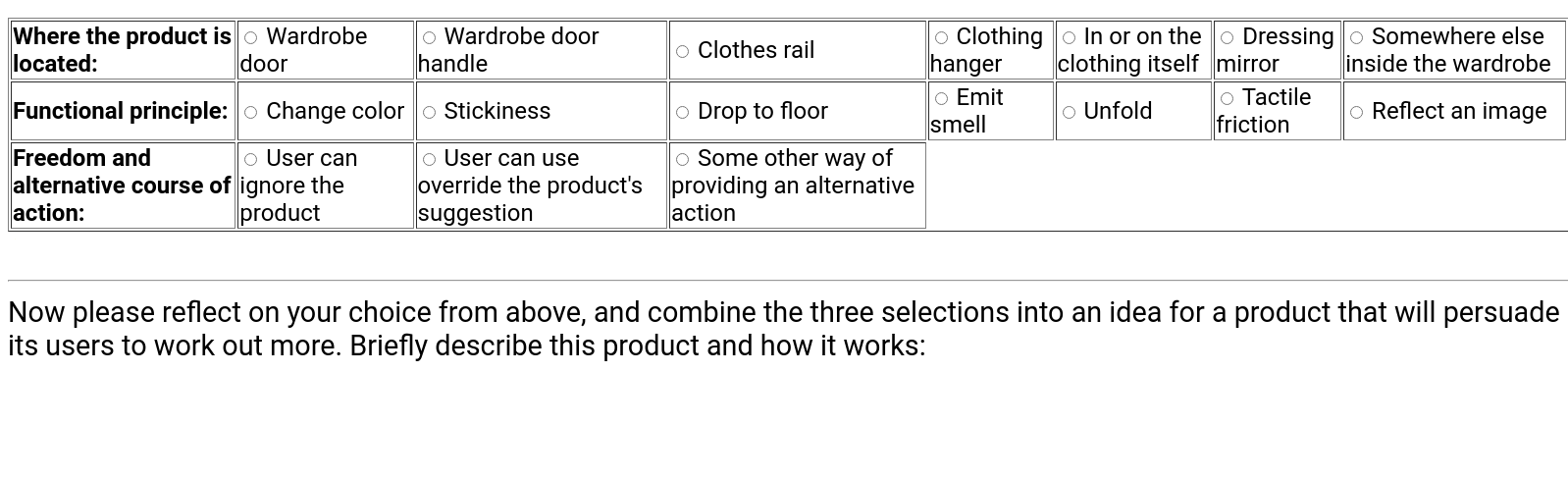}
}}%
\caption{Task given to workers in the second study.}%
\label{fig:HIT2}%
\end{figure}

To test the feasibility of this idea, we analyzed the prototyping journey described by  \citet{3461778.3462079.pdf} to curate a set of design options  for designing a persuasive product that motivates its users to physically exercise more often.
    The HIT (see Figure~\ref{fig:HIT2}) included the set of design options in a morphological matrix.
    Workers were to select one option from each row in the MM, and describe how the product would work given the three selected options.
We launched this task on MTurk using the same qualifications as above.
We coded the answers on whether 1) the worker appeared to use their cognition to solve the task, and 2) the worker provided a valid idea for a persuasive product, given the task.

\subsubsection{Study results}

Of the twelve HITs submitted by workers (9 male, 3 female, aged 22--58, mean age = 36.3 years), five (41.67\%) had to be rejected due to the same reasons as in the pilot study.
    For instance, workers entered \textit{``very useful to our quality''} and \textit{``Develop a customer centric view. A product is the item offered for sale. A product can be a service or an item,''} or some random text about the word `argument' in college writing assignments.
    Clearly, the workers tried to game the task.
Seven workers appeared to spend some cognitive effort on coming up with a solution. 
Of these seven workers, four submitted an idea for a persuasive product. 
    The four solutions are (paraphrased for brevity):%
    \begin{enumerate}%
        \item \textit{A mirror that reflects what you would look like if you continued to work out.
        The mirror includes workout plans and can be switched off to override the persuasive product.}
        \item \textit{A dressing table with a smart display.}
        \item \textit{A mirror that will swivel and provide a 360 degree view.}
        \item  \textit{Multi-colored clothing that is attractive and irresistible.}
    \end{enumerate}%
The first solution was the most elaborate and complete. 
This solution was also one of the solutions presented in the design journey by \citet{3461778.3462079.pdf}.



\section{Limitations and Future Work}%

Morphological matrices as a method for crowdsourcing ideas
require a
schematisation of the solution space into solution-independent functions.
%
    This functional structure of the end product, however, ``depends on the solutions that are considered for its implementation and vice versa'' \cite{fiorineschi2016.pdf}.
    In practice, functions are often derived by ideating temporary products.
For this reason, the approach of functional decomposition has been criticized in the literature (ibid).
Applied to the design of persuasive products, the functional solution space is vast (see \citet{3461778.3462079.pdf}).
%
This leads to two major limitations of the proposed approach. 
First, a functional and mutually-exclusive decomposition
is required to construct the MM.
Second, 
a specific goal needs to be set for the design task. For instance, instead of designing any type of product, workers would be prompted to design a product that motivates people to work out more often.

We envision a three-step procedure for applying morphological matrices in crowdsourced product design and ideation.
Workers would first generate a set of sub-problems for a given task. Workers would then define functions to address these sub-problems and determine which functions are compatible with each other. Methods from NPD, such as systematic search with the help of classification schemes \cite{pahlbeitz2007}, could be applied in this step.
This would result in a mutually-exclusive set of functions that can be assembled in a MM. 
Finally, workers would select functions from the MM 
and combine them into product ideas.
Since some functions may be mutually exclusive, the solution space could be adapted on-the-fly to the worker's selections to further facilitate the crowdsourced ideation.

\section{Conclusion}%

Morphological matrices
    hold potential to structure the idea generation process and to address the issues encountered in this paper's pilot study.
Workers in our feasibility study were able to come up with significantly more valid product ideas if their creativity was constrained with a morphological matrix.
By minimizing the number of open-ended survey items,
workers were required to select only pre-defined functions and then synthesize the selected functions into a product design.
This may have lowered the cognitive effort and lowered the creative leap in coming up with a novel product idea.

\bibliographystyle{aaai21}
\bibliography{main}

\end{document}